%% file: main.tex
\documentclass[conference]{IEEEtran}
\IEEEoverridecommandlockouts
\usepackage{amsmath,amssymb,amsfonts,amsthm}
\usepackage{comment}
\usepackage{physics}
\usepackage{glossaries}
\usepackage{hyperref}
\usepackage{lipsum} 
\usepackage{xcolor}
\usepackage{url}

\input{input.tex}
\newacronym{mimo}{MIMO}{multiple-input multiple-output}
\newacronym{simo}{SIMO}{single-input multiple-output}
\newacronym{siso}{SISO}{single-input single-output}
\newacronym{ris}{RIS}{reconfigurable intelligence surface}
\newacronym{bs}{BS}{base station}
\newacronym{ms}{MS}{mobile station}
\newacronym{csi}{CSI}{channel state information}
\newacronym{csir}{CSIR}{channel state information at the receiver}
\newacronym{tdd}{TDD}{time division duplexing}
\newacronym{los}{LOS}{line of sight}
\newacronym{ao}{AO}{alternative optimization}
\newacronym{se}{SE}{spectral efficiency}
\newacronym{aoa}{AoA}{angle of arrival}
\newacronym{nlos}{NLOS}{Non-Line-of-Sight}
\newacronym{sinr}{SINR}{signal-to-interference-plus-noise ratio}
\newacronym{snr}{SNR}{signal-to-noise ratio}
\newacronym{mmse}{MMSE}{minimum mean squared error}
\newacronym{mse}{MSE}{mean squared error}
\newacronym{ls}{LS}{least square}
\newacronym{awgn}{AWGN}{additive white Gaussian noise}
\newacronym{iid}{IID}{independent identically distributed}
\newacronym{pas}{PAS}{power angular spectrum}
\newacronym{pdpr}{PDPR}{pilot-to-data power ratio}
\newacronym{svd}{SVD}{singular value decomposition}
\newacronym{dsft}{DSFT}{discrete-spatial Fourier transform}
\newacronym{dft}{DFT}{discrete Fourier transform}
\newacronym{ue}{UE}{user equipment}
\newacronym{bf}{BF}{beamforming}

\def\txr{{\text{r}}}
\def\txb{{\text{b}}}

\def\txc{{\text{c}}}

\graphicspath{{./images/}}

\begin{document}
\title{Beamforming Saturation in Two-Timescale RIS-assisted Communication}

\bstctlcite{IEEEexample:BSTcontrol}


\author{\IEEEauthorblockN{Masoud Sadeghian$^{\star}$, Angel Lozano$^{\star}$, and Gabor Fodor$^{\dagger \sharp}$
}
\\
\normalsize\IEEEauthorblockA{$^{\star}$\emph{Department of Engineering, Universitat Pompeu Fabra, Barcelona, Spain} \\ $^{\dagger}$\emph{School of Electrical Engineering and Computer Science, KTH Royal Institute of Technology, Stockholm, Sweden}  \\ $^{\sharp}$\emph{Ericsson Research, Sweden} \\
 \{masoud.sadeghian, angel.lozano\}@upf.edu, gaborf@kth.se }
}

\maketitle
\begin{abstract}
This paper considers wireless communication assisted by a reconfigurable intelligent surface (RIS), focusing on the two-timescale approach, in which the RIS phase shifts are optimized based on channel statistics to mitigate the overheads associated with channel estimation.
It is shown that, while the power captured by the RIS scales linearly with the number of its elements, the two-timescale beamforming gain upon re-radiation towards the receiver saturates rapidly as the number of RIS elements increases, for a broad class of power angular spectra (PAS).
The ultimate achievable gain is determined by the decay rate of the PAS in the angular domain, which directly influences how rapidly spatial correlations between RIS elements diminish.  The implications of this saturation on the effectiveness of RIS-assisted communications are discussed.
\end{abstract}

\smallskip

\glsresetall 

\section{Introduction}
\label{Chapter: Intro.}

\Glspl{ris} enable creating smart radio environments in which the propagation is controlled by integrated electronic circuits and software \cite{Huang:24}.   
Since \glspl{ris} can be built by predominantly passive elements, their deployment appears
as an energy and cost efficient complement to large-scale wireless infrastructures, particularly to overcome blockages \cite{Dai:20, Tang:21}. 

Relying on passive phase-shift elements that are devoid of signal processing capabilities implies that estimating both the transmitter-\gls{ris}
and the \gls{ris}-receiver links becomes critical \cite{Long:23, Li:24}. Indeed, channel estimation in \gls{ris}-assisted wireless networks is more challenging than in conventional
systems \cite{Li:24}.
An approach to circumvent these hurdles is to optimize the phase shifts based on only the channel statistics, rather than its realizations. Sometimes referred to as a two-timescale scheme \cite{BN_two-timescale_RIS,PP_two-timescale_RIS,MX_two-timescale_RIS,MDyMS_two-timescale_RIS}, this only requires that the receiver estimates---for coherent detection purposes---the
cascade channel connecting it with the transmitter via the RIS. Put differently, the RIS becomes transparent as far as channel estimation is concerned.
This
is attractive, since standardized sophisticated channel acquisition 
procedures are supported in contemporary wireless networks.

In this paper, we consider the uplink of a \gls{ris}-aided cell in which the
\gls{bs} is equipped with multiple antennas, and ask the question of how the contribution of the RIS
depends on the number of elements it features in a two-timescale scenario.
This contribution depends on the amount of power captured by the RIS, and on its beamforming gain when re-radiating such power towards the receiver.
To address this question, an analytical expression is derived
for the \gls{snr} with optimum reception at the \gls{bs} when the \gls{ris} phase shifts are configured based on the correlation among its elements.
These correlations are determined by the topology of the RIS and by the channel's \gls{pas}.




The main result of the paper is that, for a linear RIS and
a large class of PAS functions that includes those most commonly used to model multipath propagation, the two-timescale RIS beamforming gain saturates as the number of elements grows large. 
The highest attainable gain depends on how rapidly the \gls{pas} decays angularly or, equivalently, on how rapidly the correlations decay over space.


\section{System Model}
\label{Chapter: Sys. Model}

A single-antenna \gls{ms} communicates with an $N_\text{b}$-antenna \gls{bs}, aided by a uniform linear \gls{ris} equipped with $N_\text{r}$ elements. 
Since empirical measurements indicate that the elevation angle plays a secondary role in most outdoor deployments \cite[Ch.~3]{ALRH_MIMObook}, the
RIS is horizontally disposed and the
formulation involves only the azimuth.

\subsection{Cascade Channel Model}
\label{Chapter: Channel Model}

The normalized channel between the \gls{bs} and the \gls{ris}, deterministic and of unit rank, is $\bH \in \mathbb{C}^{N_\txb \times N_\txr}$ given by
\begin{equation}
    \bH=\ba_\txb \ba_\txr^* ,
    \label{Osasuna}
\end{equation}
where $\ba_\txb$ and $\ba_\txr$ are the respective steering vectors at \gls{bs} and \gls{ris}. Precisely, $\ba_\txb$ is an arbitrary steering vector that depends on the topology of the BS and the angle of arrival thereon, while $\ba_\txr$ is a steering vector whose $m$th entry is
\begin{equation}
    [\ba_\txr]_m = \exp \! \left(\sfj 2\pi \frac{m d \cos{\theta_\txr}}{\lambda_\txc} \right) \qquad\quad m = 0, \ldots, N_\txr-1 ,
    \label{EOS}
\end{equation}
where $\theta_\txr$ is the angle of departure relative to the linear RIS, $d$ is the spacing between RIS elements, and $\lambda_\txc$ is the carrier wavelength.


The normalized channel between the \gls{ms} and the \gls{ris}, in turn, is $\bh_{\txr} \sim \mathcal{N}_\mathbb{C}(\mathbf{0},\bC_\txr)$ with covariance matrix $\bC_\txr \in \mathbb{C}^{N_\txr \times N_\txr}$.
As it models the multipath portion of the channel,  $\bC_\txr$ is nonsingular and, owing to
the linear and uniform structure of the RIS, it exhibits a Toeplitz structure; its $(k,\ell)$th entry depends only on $k - \ell = n$, being given by
\begin{equation}
    c_n = \int_{0}^{\pi} \cP(\theta) \, e^{\sfj 2 \pi  \frac{n d \cos{\theta}}{ \lambda_\txc} } \, \d \theta
    \label{eq:RIS_PAS_general}
\end{equation}
for $n=0,\ldots,N_\txr - 1$,
with $\cP(\theta)$ the \gls{pas} normalized so it can be interpreted as a probability density function \cite[Ch.~3]{ALRH_MIMObook} supported on $[0,\pi]$, which corresponds to the half-space facing the \gls{ris}. 
It follows from this PAS normalization that the entries of $\bh_{\txr}$ are of unit variance, and that $c_0 = 1$ while $c_{-n} = c^*_n$.

With the configuration of the \gls{ris} specified by the phase-shift matrix $\bPsi = \text{diag}(\bpsi)$ where $\bpsi = [e^{-j\psi_0},\dots,e^{-j\psi_{N_\txr -1}}]^\text{T}$, the cascade channel at the \gls{bs} from the \gls{ms} is
\begin{equation}
    \bh = \bH \bPsi \bh_\txr.
    \label{eq:CPA_ComChannel_t}
\end{equation}
With a statistical optimization of $\bPsi$, the cascade channel satisfies $\bh  \sim \mathcal{N}_\mathbb{C}(\mathbf{0},\bC)$ with 
\begin{align}
    \bC =  \bH \bPsi \bC_\txr \bPsi^* \bH^*, 
    \label{eq:ComCh_covariance}
\end{align}
which is singular on account of the rank-1 nature of $\bH$, and therefore of $\bh$.

While, as mentioned, the estimation of the channels to/from the RIS is utterly challenging, estimating the cascade channel $\bh$ at the BS is a standard procedure. In the case at hand, a single pilot symbol transmission suffices. By suitably repeating or power-boosting such pilot transmission, an arbitrarily precise estimate of $\bh$ can be procured \cite{MSyALyGF_RISPDPR}.
Accordingly, perfect knowledge of $\bh$ at the BS is considered in the sequel, yet the findings in the paper continue to apply in the face of channel estimation errors; they are merely obscured by the more involved formulation.

\subsection{Uplink Data Signal Model}
\label{Chapter: data signal model}

Upon uplink data transmission, the observation at the \gls{bs} is
\begin{align}
    \by = \sqrt{\alpha}  \bh x + \bn,
    \label{eq:received data}
\end{align}
where $\alpha$ is the large-scale channel gain,
$\bn \sim \mathcal{N}_\mathbb{C}(\mathbf{0},\sigma^2\bI_{N_\txb})$ is the additive white Gaussian noise, and $x$ is a data symbol of power $P$.

\section{Average SNR in Two-Timescale \\ \;\;\;\;\;\; RIS-Assisted Communication}

With $\bh$ perfectly known by the BS, the optimum receiver thereon is a linear combiner $\bw \in \mathbb{C}^{N_\txb \times 1}$ satisfying \cite{ALRH_MIMObook}
\begin{align}
    \bw \propto \bh
\end{align}
and the ensuing SNR, for a given $\bh$, equals
\begin{align}
    \text{SNR} & = \frac{\mathbb{E} \big[ \left|\bw^* \sqrt{\alpha}  \bh  x    \right|^2 | \bh \big]}{\mathbb{E} \big[ |\bw^* \bn|^2 \big]} \\
& =    \frac{\alpha P}{\sigma^2} \| \bh \|^2 .
\end{align}
Since $\bh  \sim \mathcal{N}_\mathbb{C}(\mathbf{0},\bC)$ with $\bC$ a rank-1 covariance matrix, the above SNR is exponentially distributed. Recalling \eqref{Osasuna} and (\ref{eq:ComCh_covariance}), its average value is given by
\begin{align}
 \text{SNR}_{\text{avg}} & = \frac{\alpha P}{\sigma^2} \, \mathbb{E} \big[ \| \bh \|^2 \big] \\
 & = \frac{\alpha P}{\sigma^2} \, \mathbb{E} \big[ \text{tr}(\bh \bh^*) \big] \\
 & = \frac{\alpha P}{\sigma^2} \, \tr (\bC) \\
 & = \frac{\alpha P}{\sigma^2} \, \tr (\bH \bPsi \bC_\txr \bPsi^* \bH^*) \\
 & = \frac{\alpha P}{\sigma^2} \, \tr (\bPsi \bC_\txr \bPsi^* \bH^* \bH) \\
 & = \frac{\alpha P}{\sigma^2} \, \tr (\bPsi \bC_\txr \bPsi^* \ba_\txr  \ba_\txb^* \ba_\txb \ba_\txr^*) \\
 & = \frac{\alpha P}{\sigma^2} N_\txb \, \tr (\bPsi \bC_\txr \bPsi^* \ba_\txr \ba_\txr^*) \\
 & = \frac{\alpha P}{\sigma^2} N_\txb \, \tr (\ba_\txr^* \bPsi \bC_\txr \bPsi^* \ba_\txr ) \\
 & = \frac{\alpha P}{\sigma^2} N_\txb \, \ba_\txr^* \bPsi \bC_\txr \bPsi^* \ba_\txr \\
 & = \frac{\alpha P}{\sigma^2} N_\txb N_\txr \, \bv^* \bPsi \bC_\txr \bPsi^* \bv \label{EOS1} \\
 & = \frac{\alpha P}{\sigma^2} N_\txb \!\!\!\!\!\!\!\!\! \underbrace{ N_\txr \, \zeta}_{\text{RIS power gain}} \!\!\!\!\!\!\!\!   \label{EOS2}
\end{align}
where in \eqref{EOS1} and \eqref{EOS2} we introduced, respectively, 
\begin{align}
\bv= \frac{1}{\sqrt{N_\txr}}  \ba_\txr
\end{align}
and
\begin{align}
\zeta & =\bv^* \bPsi \bC_\txr \bPsi^* \bv \\
& = \bpsi \, \text{diag}(\bv^*) \bC_\txr \text{diag}(\bv) \bpsi^* .
\label{UoM}
\end{align}
Within the average SNR in \eqref{EOS2}, the multiplier $N_\txb$ corresponds to the gain of the BS receive array, which leaves the product $N_\txr \zeta$ as the power gain provided by the RIS. This gain is made up of two factors:
\begin{itemize}
    \item $N_\txr$, which reflects the power captured by the RIS acting as a receiver and does not depend on the RIS phase shifts,
    \item $\zeta$, which represents the beamforming gain provided by the RIS acting as a transmitter, incorporating the effect of the RIS phase shifts. One can view $\zeta$ as the average power gain relative to these phase shifts being random.
\end{itemize}
If the channel realizations to/from the RIS could be perfectly estimated, the RIS phase shifts could be instantaneously optimized yielding $\zeta = N_\txr$, for a RIS power gain $N_\txr \zeta$ that would altogether be quadratic in the number of RIS elements \cite{WQyRZ_QuadraticPower}. 
In a two-timescale system, however, the beamforming gain $\zeta$ behaves rather differently.


\section{Two-Timescale Beamforming Gain}
\label{Chapter: RIS}

The optimum phase shifts are those that maximize the beamforming gain, $\zeta$.
The optimization of the RIS phase shifts can thus be cast, recalling (\ref{UoM}), as
\begin{align}
    \zeta & = \max_{\bpsi: \, |\psi_m|=1 \; \forall m}  \bpsi \, \text{diag}(\bv^*) \bC_\txr \text{diag}(\bv) \, \bpsi^* 
\label{eq:RIS_opt}
\end{align}
whose objective function is quadratic, but whose unit-magnitude constraints are not convex.
This problem is addressed in \cite{AM_SDP_QO} with an approximation guarantee, while \cite{ASyPD_RISoptimization} introduces a low-complexity alternative optimization approach.
Sidestepping numerical optimization procedures, in this section insights are gleaned analytically.

For $N_\txr=1$, trivially $\zeta=1$ in a two-timescale system as it would be with instantaneously optimized phase shifts. For $N_\txr=2$, however, the latter would double to $\zeta=2$ while the former amounts, after a bit of algebra, to 
\begin{align}
    \zeta & = \max_{\psi_0,\psi_1: \, |\psi_0|=1, |\psi_1|=1} 1 + \Re \! \left \{ c_1 e^{-\sfj (\psi_1 - \psi_0)} e^{\sfj 2 \pi  \frac{d \cos{\theta}}{ \lambda_\txc} } \right \} ,
\end{align}
which, by setting $\psi_1 - \psi_0 = \arg(c_1)+ 2 \pi d \cos \theta_\txr / \lambda_\txc$, returns
\begin{align}
    \zeta = 1 + |c_1| .
\end{align}
As $\bC_\txr$ is nonsingular, $|c_1|<1$ and thus the two-timescale beamforming gain satisfies $\zeta < 2$, falling short of its instantaneously optimized counterpart.
In fact, the two-timescale beamforming gain is strictly sublinear in $N_\txr$ for every $N_\txr$. This can be appreciated by relaxing the constraints in \eqref{eq:RIS_opt} to $\|\bpsi\| =\sqrt{N_\txr}$, whereby
\begin{align}
    \zeta & \leq \max_{\bpsi: \, \|\bpsi\| =\sqrt{N_\txr} }  \bpsi \, \text{diag}(\bv^*) \bC_\txr \text{diag}(\bv) \, \bpsi^* \\
    & =  N_\txr \, \lambda_\text{max} \! \Big( \text{diag}(\bv^*) \bC_\txr \text{diag}(\bv) \Big) ,
    \label{Fermin}
\end{align}
where $\lambda_\text{max}(\cdot)$ denotes the largest eigenvalue of a matrix, and (\ref{Fermin}) is achieved by the maximum-eigenvalue eigenvector of $\text{diag}(\bv^*) \bC_\txr \text{diag}(\bv)$.
Since $\sqrt{N_\txr} \, \text{diag}(\bv^*) $ is a unitary matrix, the above is tantamount to
\begin{align}
    \label{eq:relaxMax}
    \zeta & \leq \lambda_\text{max} (  \bC_\txr ) \\
    & < \text{tr}(\bC_\txr) \label{Olmo} \\
    & = N_\txr ,
\end{align}
with (\ref{Olmo}) following from the nonsingular condition of $\bC_\txr$.

If the RIS elements are uncorrelated, then $\lambda_{\text{max}}(\bC_\txr) = 1$ and
the above corroborates the stronger result that, as expected, there is no beamforming gain to be had in a two-timescale system without RIS element correlation. In the sequel, therefore, only situations in which the RIS elements do exhibit correlation are considered.

Owing to the structure of $\bv$, which descends from \eqref{EOS}, the $(k,\ell)$th entry of  $\text{diag}(\bv^*) \bC_\txr \text{diag}(\bv)$ equals
\begin{align}
\left[ \text{diag}(\bv^*) \bC_\txr \text{diag}(\bv) \right]_{k,\ell} = c_n \; \frac{ e^{\sfj 2 \pi n d \cos \theta_\txr / \lambda_\txc} }{N_\txr} ,
\end{align}
where, recall, $n=k - \ell$. Thus, the Toeplitz structure of $\bC_\txr$ is readily inherited by
$\text{diag}(\bv^*) \bC_\txr \text{diag}(\bv)$, whereby 
its maximum eigenvalue is upper bounded by $ \sum_{n=-\infty}^{\infty}  |c_n|$ \cite{gray_toeplitz2006},
such that
\begin{align}
    \zeta & \leq \!\! \sum_{n=-\infty}^{\infty}  |c_n| .
    \label{regio7}
\end{align}
The beamforming gain is thus sure to be curbed if the right-hand side of (\ref{regio7}) is finite, which corresponds to $\bC_\txr$ being a Wiener-class Toeplitz matrix.
This, in turn, is guaranteed if
\begin{equation}
|c_n| = o \!  \left( \frac{1}{|n|} \right),
\label{ALS}
\end{equation}
which, as seen in Sec. \ref{examples}, is satisfied by a number of functions that have been shown to be a satisfactory fit for the PAS.

Irrespective of whether \eqref{ALS} holds, a  precise assertion of how the beamforming gain behaves can be made asymptotically in the number of \gls{ris} elements.
Indeed, for $N_\txr \to \infty$, numerical optimizations become unnecessary as
(\ref{eq:RIS_opt}) admits a closed-form solution. 
Under mild conditions satisfied by any well-behaved PAS, for growing $N_\txr$ the eigenvectors of $\text{diag}(\bv^*) \bC_\txr \text{diag}(\bv)$ converge, on account of the Toeplitz nature of this matrix, to Fourier vectors 
of the form \cite{gray_toeplitz2006}
\begin{align}
    \frac{1}{\sqrt{N_\txr}} \left[1,e^{-\sfj 2\pi  m / N_\txr},\ldots,e^{- \sfj 2\pi m (N_\txr-1) / N_\txr} \right]^*
    \label{eq:eigDFT}
\end{align}
for $m= 0,..., N_\txr-1$.
As these vectors (scaled by $N_\txr$) comply with the constraints in \eqref{eq:RIS_opt}, the solution to this original problem is sure to equal that of the relaxed problem in \eqref{eq:relaxMax}, meaning that the optimum beamforming gain satisfies
\begin{align}
    \zeta & \approx
      \lambda_\text{max} (  \bC_\txr ),
      \label{coffee}
\end{align}
which sharpens with growing $N_\txr$ and becomes exact asymptotically.
The RIS phase shifts achieving this beamforming gain, in turn, converge to the Fourier eigenvector (scaled by $\sqrt{N_\txr}$) associated with the maximum eigenvalue of $\bC_\txr $. As seen next, for many PAS functions of interest this eigenvalue is bounded from above.

\section{Examples}
\label{examples}

Let us next entertain some of the functions most commonly considered to model the PAS. 

\begin{itemize}
    
    \item \label{truncatedGaussian_pas} \textbf{Truncated Gaussian}. 
    A Gaussian function with mean $\mu_\theta$ and angular spread $\sigma_\theta > 0$, whose support is restricted to $\theta \in (0,\pi)$, has been shown to offer a good fit to the empirical PAS in elevated \glspl{bs} \cite{Chu_TrunGauPAS}. The function is 
    \begin{equation}
        \cP(\theta)=\frac{K_\sfG}{\sqrt{2}\sigma_\theta} \exp\left(-\frac{(\theta-\mu_\theta)^2}{2 \sigma_\theta^2}\right),
    \end{equation}
    where
    \begin{equation}
        K_\sfG = \frac{1}{\sqrt{\pi}[1-2 Q(\frac{\pi}{2\sigma_\theta})]},
    \end{equation}
    is a normalization factor ensuring that the PAS integrates to one, and $Q(\cdot)$ is the Gaussian Q-function.
    
    For small $\sigma_\theta$, the integration in \eqref{eq:RIS_PAS_general} yields \cite[Ch.~3]{ALRH_MIMObook}
    \begin{equation}
        |c_n| \approx \exp \! \left( -2 \left[ \frac{\pi d n \sin(\mu_\theta) \sigma_\theta}{\lambda_\txc} \right]^2 \right),
        \label{eq:TrunGauPas}
    \end{equation}
   whereby (\ref{regio7}) specializes to \cite[Ch.~16.27]{abramowitz+stegun}
     \begin{equation}
       \zeta \leq \vartheta \! \left(0 , e^{-2 \left[ \frac{\pi d \sin(\mu_\theta) \sigma_\theta}{\lambda_\txc} \right]^2} \right),
        \label{eq:GauPasOpt}
    \end{equation}
    where $\vartheta(\cdot,\cdot)$ is the Jacobi theta function. 
    The decay of \eqref{eq:TrunGauPas} is decidedly faster than $1/|n|$ and the beamforming gain is therefore curbed.
   

    \item \label{truncatedLaplacian_pas} \textbf{Truncated Laplacian}.
    Extensive measurements have shown a strong fit of the Laplacian \gls{pas}, with a properly tuned $\sigma_\theta$ and support on $(0,\pi)$, to both indoor \cite{Swind_LaplacianIndoor} and outdoor environments \cite{KPyBF_LaplacianPAS}. This function is  \cite[Ch.~3]{ALRH_MIMObook} 
    \begin{equation}
        \cP(\theta)=\frac{K_\sfL}{\sqrt{2}\sigma_\theta} \exp\left(-\left| \frac{\sqrt{2}(\theta-\mu_\theta)}{\sigma_\theta} \right| \right),
    \end{equation}
    with
    \begin{equation}
        K_\sfL = \frac{1}{1- \exp \! \left(\frac{\pi}{\sqrt{2}\sigma_\theta}\right)} .
    \end{equation}
    For small $\sigma_\theta$, it holds that   \cite{RH_LaplacianPAS}
    \begin{equation}
        |c_n| \approx \frac{1}{1+2(\pi d n \sin(\mu_\theta)\sigma_\theta/\lambda_\txc)^2} 
        \label{eq:TrunLapPas}
    \end{equation}
    and, from
    \begin{equation}
        \pi \cot(\pi z) = \frac{1}{z} + 2z\sum_{n=1}^{\infty} \frac{1}{z^2-n^2},
    \end{equation}
    it follows that (\ref{regio7}) specializes to
    \begin{equation}
        \zeta \leq \frac{\lambda_\txc}{\sqrt{2} d \sin(\mu_\theta)  \sigma_\theta } \coth \! \left( \frac{\lambda_\txc}{\sqrt{2} d \sin(\mu_\theta)  \sigma_\theta } \right). 
        \label{eq:LapPasOpt}
    \end{equation}
    In this case, $|c_n| = \cO(1/|n|^2)$, whereby the beamforming gain is again curbed. 

    \item \label{Exponential_pas} \textbf{Exponential correlation model}.
    This simpler model is
    \begin{equation}
        |c_n|=\kappa^{|n|},
    \end{equation}
    where $\kappa < 1$ is a parameter \cite{Loyka_ExpPAS}.
    Leveraging the properties of the geometric series, \eqref{regio7} specializes to
    \begin{equation}
        \zeta \leq \frac{1 + |\kappa|}{1 - |\kappa|} ,
        \label{eq:ExpPasOpt}
    \end{equation}
    which is always bounded.

\end{itemize}

\begin{figure}[t]
    \centering 
    \includegraphics[width=.999\linewidth]{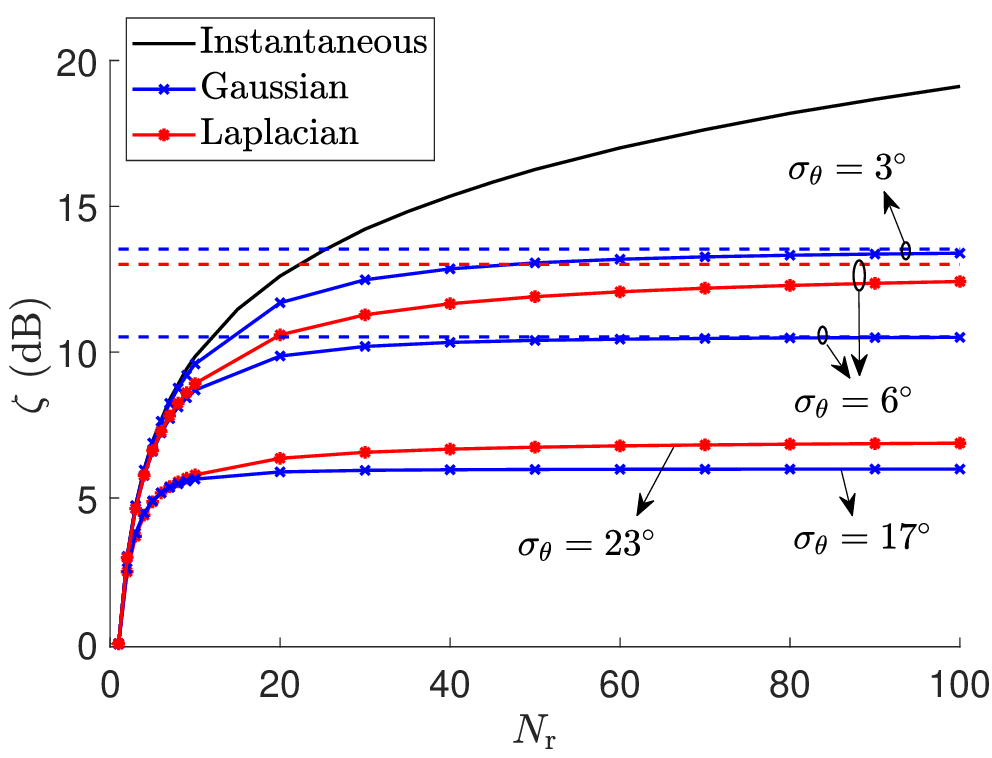}
    \vspace{-6mm}
    \caption{Beamforming gain as a function of the number of \gls{ris} elements for different \glspl{pas}, with the \gls{ris} element spacing set to $d=\lambda_\txc/2$. For the truncated Gaussian \gls{pas}, $\mu_\theta=\pi/4$ and $\sigma_\theta=\{3^\circ,6^\circ,17^\circ\}$ \cite{Chu_TrunGauPAS}. For the truncated Laplacian \gls{pas}, $\mu_\theta=\pi/4$ and $\sigma_\theta=\{6^\circ,23^\circ\}$ \cite{RH_LaplacianPAS,Swind_LaplacianIndoor}. Included as benchmark is the gain with instantaneously optimized phase shifts.}
    \label{fig:lineararray_maxeig}
\end{figure}

The beamforming gain for the truncated Gaussian and Laplacian PAS functions is exemplified in
Fig. \ref{fig:lineararray_maxeig} as a function of the number of \gls{ris} elements, with representative angular spreads and numerically computed RIS phase shifts.
The dashed lines indicate the upper bounds to the beamforming gains, for those cases in which expressions are available; these upper bounds are seen to be tight, and they are approached rapidly as $N_\txr$ grows large. Also included in the figure is the beamforming gain obtainable with instantaneously optimized RIS phase shifts, i.e., $\max_{\bPsi} \bbE \! \left[ \bv^* \bPsi \bh_\txr \bh_\txr^* \bPsi^* \bv\right] $; this gain grows linearly and sustainedly with $N_\txr$.

\begin{figure}[t]
    \centering
    \includegraphics[width=.999\linewidth]{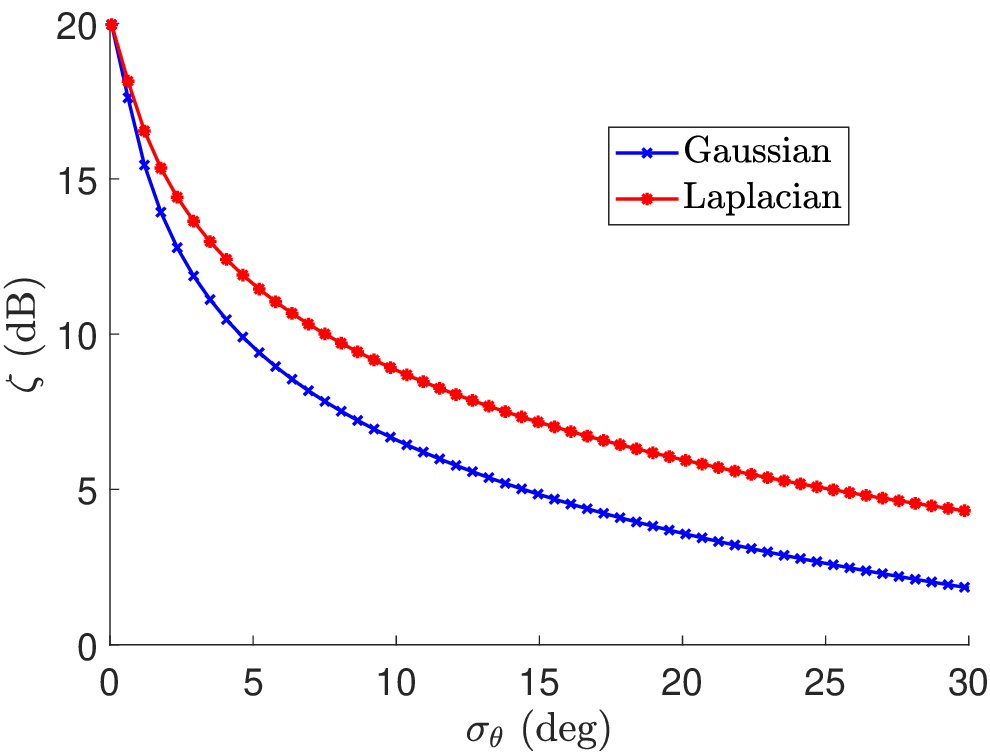}
    \vspace{-6mm}
    \caption{Beamforming gain as a function of the angular spread for truncated Gaussian and Laplacian PAS, with $N_\txr=100$, $\mu_\theta=\pi/2$, and $d=\lambda_\txc/2$.}
    \vspace{-2.5mm}
    \label{fig:AngularSpread}
\end{figure}

To complement the above, Fig. \ref{fig:AngularSpread} depicts the beamforming gain as a function of the angular spread, with $N_\txr=100$ as a proxy for $N_\txr \to \infty$.
Not only is the gain curbed, but it is restricted to values that are modest and that decay rapidly. Unless $\sigma_\theta < 10^\circ$, we have that $\zeta < 8$~dB for both the Gaussian and the Laplacian PAS, no matter how large the RIS may be.
For $\sigma_\theta \to 0$, the MS–RIS channel becomes line-of-sight and $\bC_\txr$ becomes rank-$1$, whereby the beamforming gain grows unboundedly with the number of RIS elements.


\begin{figure}[t]
    \centering
    \includegraphics[width=.99\linewidth]{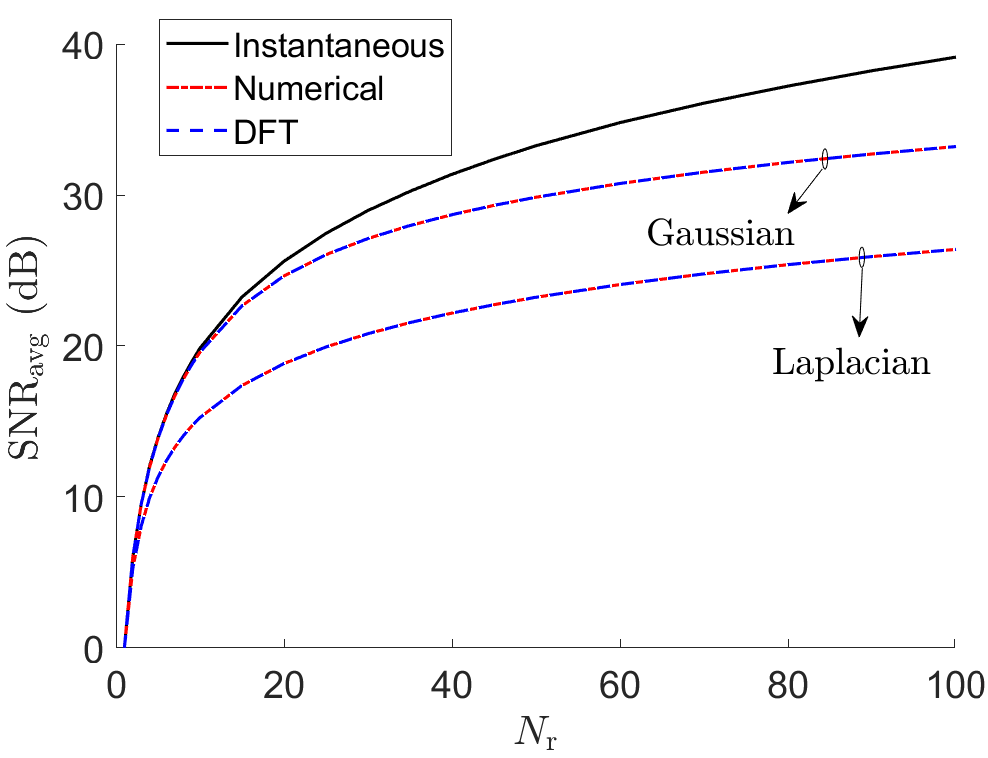}
    \vspace{-6mm}
    \caption{Average SNR versus $N_{\txr}$ with $N_\txb=10$, $\theta_\txr=80^{\circ}$, $\mu_\theta = \pi/4$, and ${\alpha P}/{\sigma^2}=-10$~dB. 
    For the truncated Gaussian \gls{pas}, $\sigma_\theta=3^\circ$ \cite{Chu_TrunGauPAS}. For the truncated Laplacian \gls{pas}, $\sigma_\theta = 23^\circ$ \cite{Swind_LaplacianIndoor}.
    Results shown for phase shifts computed numerically and for their asymptotic DFT values.
    Included as benchmark is the gain with instantaneously optimized phase shifts.
    } 
    \vspace{-3mm}
    \label{fig:SNR}
\end{figure}

The average SNRs corresponding to some of the curves in Fig. \ref{fig:lineararray_maxeig} are depicted in Fig. \ref{fig:SNR}.
As the number of \gls{ris} elements increases, instantaneous phase-shift tuning delivers a quadratic improvement with $N_\txr$ while, in the two-timescale examples, the asymptotic improvement is merely linear.
Also noteworthy in the two-timescale examples is that the SNRs with
the RIS phase shifts optimized numerically as per \cite{MSyALyGF_RISPDPR} 
are barely distinguishable from those with the asymptotically optimum Fourier values in \eqref{eq:eigDFT}. This excellent match indicates that the Fourier phase shifts, of lower computational complexity, are an enticing solution to configure a two-timescale linear RIS even for small $N_\txr$.
%



\section{Conclusion}
\label{Chapter: Conclusion} 

The two-timescale approach to RIS-assisted communication avoids the costly estimation of the channels to/from the RIS. Moreover, computationally attractive Fourier phase shifts can be applied in the case of a linear RIS. This overall convenience, however, comes at a price in performance. The power gain, quadratic in the number of RIS elements if
the RIS phase shifts could be instantaneously optimized, reverts to a scaling that is always sub-quadratic; for a broad class of channels, in fact, this scaling becomes merely linear beyond a relatively small number of elements.
This poorer scaling 
with the number of elements curtails the ability of the RIS to provide a decisive advantage relative to ambient propagation \cite{MAA_indoorRIS,Reinaldo_RIS}, and renders the RIS less competitive against alternatives such as relays and  network-controlled repeaters \cite{EByEL_RISvsRelays}. 
%
The generalization of these findings to planar RIS entails dealing with block-Toeplitz covariance matrices. The results are qualitatively similar, but a larger number of RIS elements is required for the power gain scaling to become merely linear.



\section*{Acknowledgment}
\vspace{-0.5mm}

Work supported by the Horizon 2020 MSCA-ITN-METAWIRELESS Grant Agreement 956256, by
MICIU under grant CEX2021001195-M, by ICREA, and by the  Swedish SSF project SAICOM Grant No: FUS21-0004.



{
\small
\bibliographystyle{IEEEtran}
\bibliography{refs}
}
\end{document}

%% file: input.tex
\usepackage[numbers,sort&compress]{natbib} 

\usepackage{amsmath}
\usepackage{numprint} 
\usepackage{amsthm} 
\usepackage{url} 
\usepackage{textcomp} 

\usepackage{accents} 
\usepackage{amssymb}
\usepackage{amsfonts}
\usepackage{dsfont}
\usepackage[figuresright]{rotating} 
\usepackage{floatpag} 
	\rotfloatpagestyle{empty} 

\usepackage{algorithm}
\usepackage{algpseudocode}
\usepackage{longtable}
\usepackage{subfigure}

\usepackage{epstopdf} 
\usepackage{makeidx} 
\usepackage{upgreek} 
\makeindex 

\usepackage{times}
\usepackage{mathdots} 

\usepackage{bm}
\usepackage{bbm}

\theoremstyle{plain}

\def\bb0{{\mathbb{0}}}

\def\ba{{\boldsymbol{a}}}
\def\bb{{\boldsymbol{b}}}

\def\bh{{\boldsymbol{h}}}

\def\bn{{\boldsymbol{n}}}

\def\bv{{\boldsymbol{v}}}
\def\bw{{\boldsymbol{w}}}

\def\by{{\boldsymbol{y}}}

\def\b0{{\boldsymbol{0}}}

\def\bC{{\boldsymbol{C}}}

\def\bH{{\boldsymbol{H}}}
\def\bI{{\boldsymbol{I}}}

\def\b{{\mathrm{b}}}

\def\d{{\mathrm{d}}}

\def\r0{{\mathbf{0}}}

\def\bbE{{\mathbb{E}}}

\def\cO{\mathcal{O}}
\def\cP{\mathcal{P}}

\def\sfG{\mathsf{G}}

\def\sfL{\mathsf{L}}

\def\bpsi{\bm \psi}

\def\bPsi{\bm \Psi}

\def\sfj{{\mathsf{j}}}

\def\bsf0{{\bm{\mathsf{0}}}}

\def\N0{{N_{\mathrm{0}}}}

\def\bsf{{\boldsymbol{s}_\mathrm{f}}}